\documentclass[aps,pra,reprint,superscriptaddress,showpacs]{revtex4-1}
\usepackage{graphicx,amsmath,amsfonts,natbib}
\newcommand{\ket}[1]{|#1\rangle}

\begin{document}

\title{Spectroscopy on a single trapped $^{137}$Ba$^+$ ion \\for nuclear magnetic octupole moment determination}

\author{Nicholas C. Lewty}
\email{nicholas.lewty@nus.edu.sg}
\affiliation{Centre for Quantum Technologies and Department of Physics, National University of Singapore, 3 Science Drive 2, 117543 Singapore}
\author{Boon Leng Chuah}
\affiliation{Centre for Quantum Technologies and Department of Physics, National University of Singapore, 3 Science Drive 2, 117543 Singapore}
\author{Radu Cazan}
\affiliation{Centre for Quantum Technologies and Department of Physics, National University of Singapore, 3 Science Drive 2, 117543 Singapore}
\author{B. K. Sahoo}
\affiliation{Theoretical Physics Division, Physical Research Laboratory, Ahmedabad-380009, India}
\author{M. D. Barrett}
\affiliation{Centre for Quantum Technologies and Department of Physics, National University of Singapore, 3 Science Drive 2, 117543 Singapore}

\date{\today}

\begin{abstract}
We present precision measurements of the hyperfine intervals in the $5\mathrm{D}_{3/2}$ manifold of a single trapped Barium ion, $^{137}\mathrm{Ba}^+$. Measurements of the hyperfine intervals are made between $m_F=0$ sublevels over a range of magnetic fields allowing us to interpolate to the zero field values with an accuracy below a few Hz, an improvement on previous measurements by three orders of magnitude. Our results, in conjunction with theoretical calculations, provide a 30-fold reduction in the uncertainty of the magnetic dipole ($A$) and electric quadrupole ($B$) hyperfine constants. In addition, we obtain the magnetic octupole constant ($C$) with an accuracy below 0.1~Hz. This gives a subsequent determination of the nuclear magnetic octupole moment, $\Omega$, with an uncertainty of $1\%$ limited almost completely by the accuracy of theoretical calculations. This constitutes the first observation of the octupole moment in $^{137}\mathrm{Ba}^+$ and the most accurately determined octupole moment to date.
\end{abstract}


\maketitle
High precision measurements of the hyperfine structure provides stringent tests for state-of-the-art atomic structure calculations. These calculations play a crucial role in the interpretation of parity nonconservation (PNC) experiments which provide important tests of the standard model at low energy \cite{Langacker_RMP}. The accuracy of calculated PNC matrix elements can be assessed by comparing measured hyperfine structure constants with calculated values \cite{Ba+_pnc}. In addition, the hyperfine structure provides insight into the nuclear structure of atoms \cite{Arimondo}.

In this paper we present precision measurements of the hyperfine intervals of the 5D$_{3/2}$ manifold of $^{137}\mathrm{Ba}^+$. By combining high precision radio frequency (rf) spectroscopy with shelving techniques \cite{Dietrich,Boon Leng} on singly trapped ions, we measure the hyperfine intervals of the 5D$_{3/2}$ manifold to an accuracy below a few Hz. Together with theoretical calculations, this permits a 30-fold reduction in the uncertainty of the magnetic dipole ($A$) and electric quadrupole ($B$) hyperfine constants and the first observation of the magnetic octupole moment in $^{137}\mathrm{Ba}^+$. We determine the magnetic octupole moment, $\Omega$, to an accuracy of $1\%$ percent limited almost entirely by uncertainties in the theory.

Ba$^+$ is an excellent candidate for magnetic octupole determination. Long lived metastable $D$ states permit high precision spectroscopy measurements of the hyperfine levels, and high precision calculations are possible~\cite{Beloy_pra_012512_2008}. In \cite{Beloy_pra_052503_2008} it was shown that the hyperfine intervals, $\delta W_F = W_F-W_{F+1}$, of the 5D$_{3/2}$ manifold can be written
\begin{equation}
\delta W_0=-A+B-56C-\frac{1}{100}\eta+\frac{\zeta}{100}\sqrt{\frac{7}{3}},
\end{equation}
\begin{equation}
\delta W_1=-2A+B+28C-\frac{1}{75}\eta,
\end{equation}
\begin{equation}
\delta W_2=-3A-B-8C-\frac{1}{300}\eta-\frac{\zeta}{20}\sqrt{\frac{3}{7}},
\end{equation}
where $C$ is the magnetic octupole hyperfine constant while $\eta$ and $\zeta$ are the correction terms characterizing the mixing with the upper 5D$_{5/2}$ manifold. With measured values of the hyperfine intervals, $\delta W_k$, and theoretical estimates for the correction terms we can solve the equations for the hyperfine constants. The octupole moment, $\Omega$, can then be extracted from our theoretical result \cite{Sahoo_prep}
\begin{equation}
\label{omega}
C=-0.581(6)\left( \frac{\Omega}{\mu_N \times \rm{b} }\right)~\rm{kHz},
\end{equation}
using the CCSD(T) method described in \cite{Ba+_pnc,bijaya_ba+_atom_quad}, where $\mu_N$ is the Bohr magneton and b is the barn unit of area.

\begin{figure}
\includegraphics[width=0.48\textwidth]{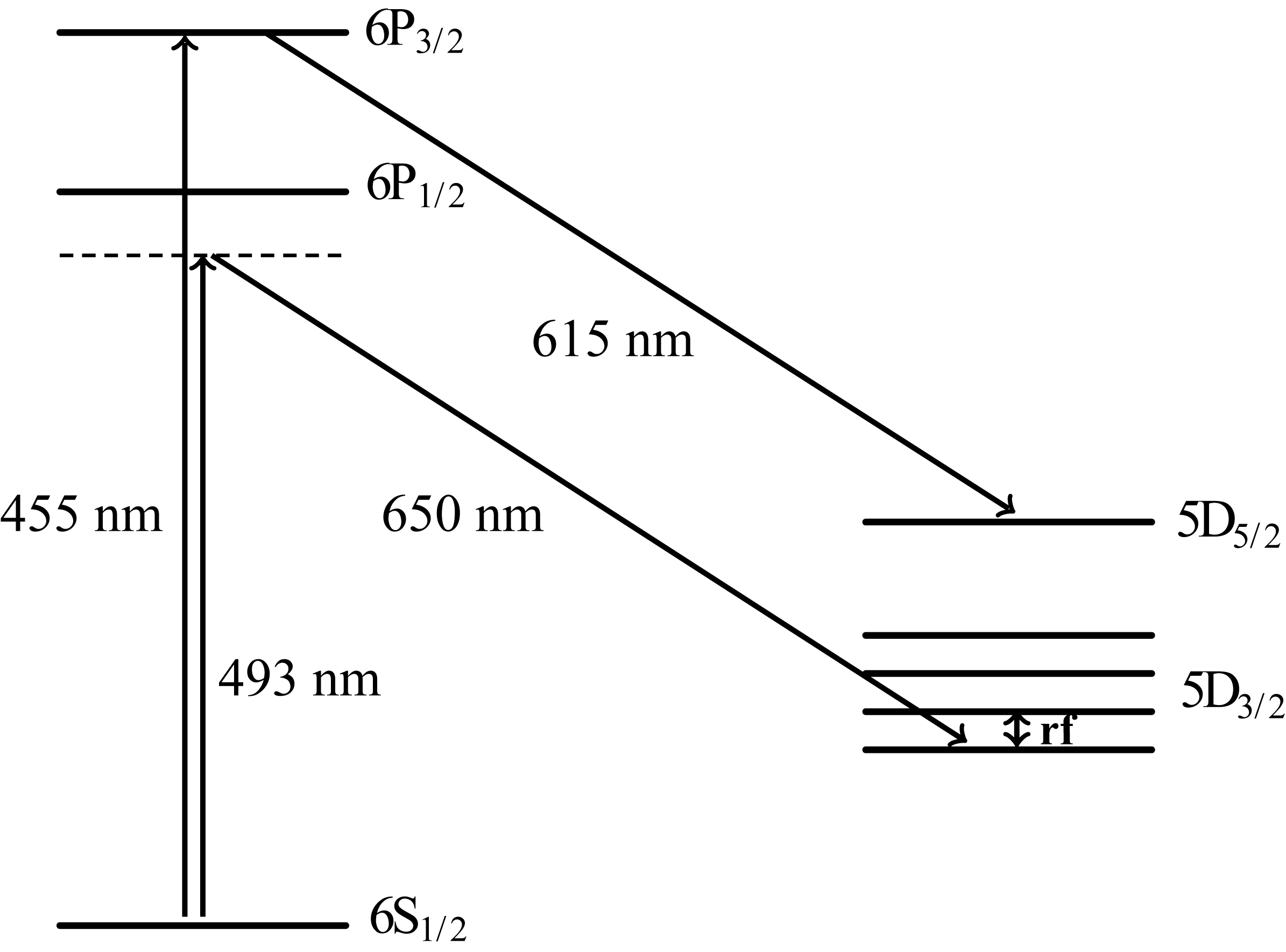}
\caption{Relevant levels of $^{137}$Ba$^{+}$ for the rf spectroscopy: The ion is prepared in the $\left|F=2, m_F=2\right\rangle$ state of the 6S$_{1/2}$ level from where it is shelved to $\left|F'', m_{F''}=0\right\rangle$ of the 5D$_{3/2}$ level with a pair of Raman beams which are red-detuned by $\approx \Delta=2\pi\times 500\,\mathrm{GHz}$ from the 6P$_{1/2}$ level. The rf transition is detected by shelving to the 5D$_{5/2}$ level (see text).}
\label{Ba_levels}
\end{figure}

The procedure to measure the hyperfine intervals is similar to that proposed in \cite{koerber_prl_2002,Dietrich} and the relevant level structure is given in Fig.~\ref{Ba_levels}. The ion is first Doppler cooled and then optically pumped into the state $\ket{F=2,m_F=2}$ of the 6S$_{1/2}$ level. To measure the hyperfine interval, $\delta W_k$, the ion is then shelved to the state $\ket{F''=k,m_F=0}$ of the 5D$_{3/2}$ level using a two photon Raman transition similar to the one used in \cite{Boon Leng}. An rf antenna is then turned on to drive $F''\leftrightarrow F''+1$ transitions. The signal generator used for the rf antenna is synchronized with a GPS-disciplined Rb clock.

To determine if the hyperfine transition occurred we use a second Raman pulse to transfer the state $\ket{F''=k,m_F=0}$ back to the $\ket{F=2,m_F=2}$ ground state and then optically pump on the 6S$_{1/2}$ to 6P$_{3/2}$ transition at 455~nm. If the hyperfine transition occurred, the ion will remain in the $\ket{F''=k+1,m_F=0}$ state of the 5D$_{3/2}$ level, otherwise it will be shelved to the 5D$_{5/2}$ level with $\approx 88\%$ efficiency. Subsequent driving of the 6S$_{1/2}$ to 6P$_{1/2}$ and 5D$_{3/2}$ to 6P$_{1/2}$ transitions using the Doppler cooling beams then provides a fluorescence measurement for the probability of driving the hyperfine transition: the ion being bright if the hyperfine transition took place and dark otherwise. The detection efficiency of this scheme is limited to approximately $88\%$ by the branching ratio between the 6P$_{3/2}$ and 5D$_{3/2}$ which results in unwanted population of the 5D$_{3/2}$ level when optically pumping to the 5D$_{5/2}$ level. This does not impact on the accuracy at which we can measure the hyperfine transition probability, but only on the amount of averaging needed to achieve a particular level of accuracy.

The experiments are performed in a four-rod linear Paul trap, similar to the ones described in \cite{Berkeland,Boon Leng}. The trap consists of four stainless steel rods of diameter 0.45~mm whose centers are arranged on the vertices of a square with 2 mm length of the side. A 5.3~MHz rf potential with an amplitude of 170~V is applied via a step-up transformer to two diagonally opposing electrodes. A small DC voltage applied to the other two electrodes ensures a splitting of the transverse trapping frequencies and rotates the principle axes of the trap with respect to the propagation direction of the cooling lasers. Axial confinement is provided by two axial needle electrodes separated by 2.4~mm and held at 33~V. Using this configuration, the measured trapping frequencies are $(\omega_x, \omega_y, \omega_z)/2\pi \approx (1.7, 1.5, 0.5)$~MHz.

Doppler cooling and detection is achieved by driving the 6S$_{1/2}$ to 6P$_{1/2}$ transitions at $493$~nm and repumping on the 5D$_{3/2}$ to 6P$_{1/2}$ transitions at $650$~nm. The $493$~nm laser is passed through two electro-optic modulators in order to generate the sidebands required to address all possible transitions between the 6S$_{1/2}$ to 6P$_{1/2}$ levels. Similarly, to address all the 5D$_{3/2}$ to 6P$_{1/2}$ transitions, the repumping laser at 650~nm is split into four, frequency shifted by acousto-optic modulators, and then recombined into a single fiber. All laser fields are linearly polarized perpendicular to a magnetic field of approximately 1~G. This configuration avoids unwanted dark states in both the cooling and detection cycles. Additionally, the field ensures a well defined quantization axis for optical pumping and state preparation.

The magnetic field also gives a second order Zeeman shift of the $\ket{F'',m_F=0}$ levels which must be accounted for when measuring the hyperfine intervals. We achieve this by additionally measuring transitions between $\ket{F'',m_F=0}$ and $\ket{F''+1,m_F=\pm 1}$. Half of the difference frequency between the $\Delta m_F=+1$ and the $\Delta m_F=-1$ transitions is then given by the linear shift $\mu_B g_F B/\hbar$ with any quadratic shifts cancelled. By measuring the $\Delta m_F=0$ transition over a range of magnetic fields, each calibrated by measurements of the $\Delta m_F=\pm 1$ transitions, we can map out the full field dependence of the $\Delta m_F=0$ transition and interpolate to the desired zero field result. This also yields the second order Zeeman shift coefficient for the $\Delta m_F=0$ transitions providing a useful consistency check within our measurements.

\begin{figure}[htb]
\includegraphics[width=0.48\textwidth]{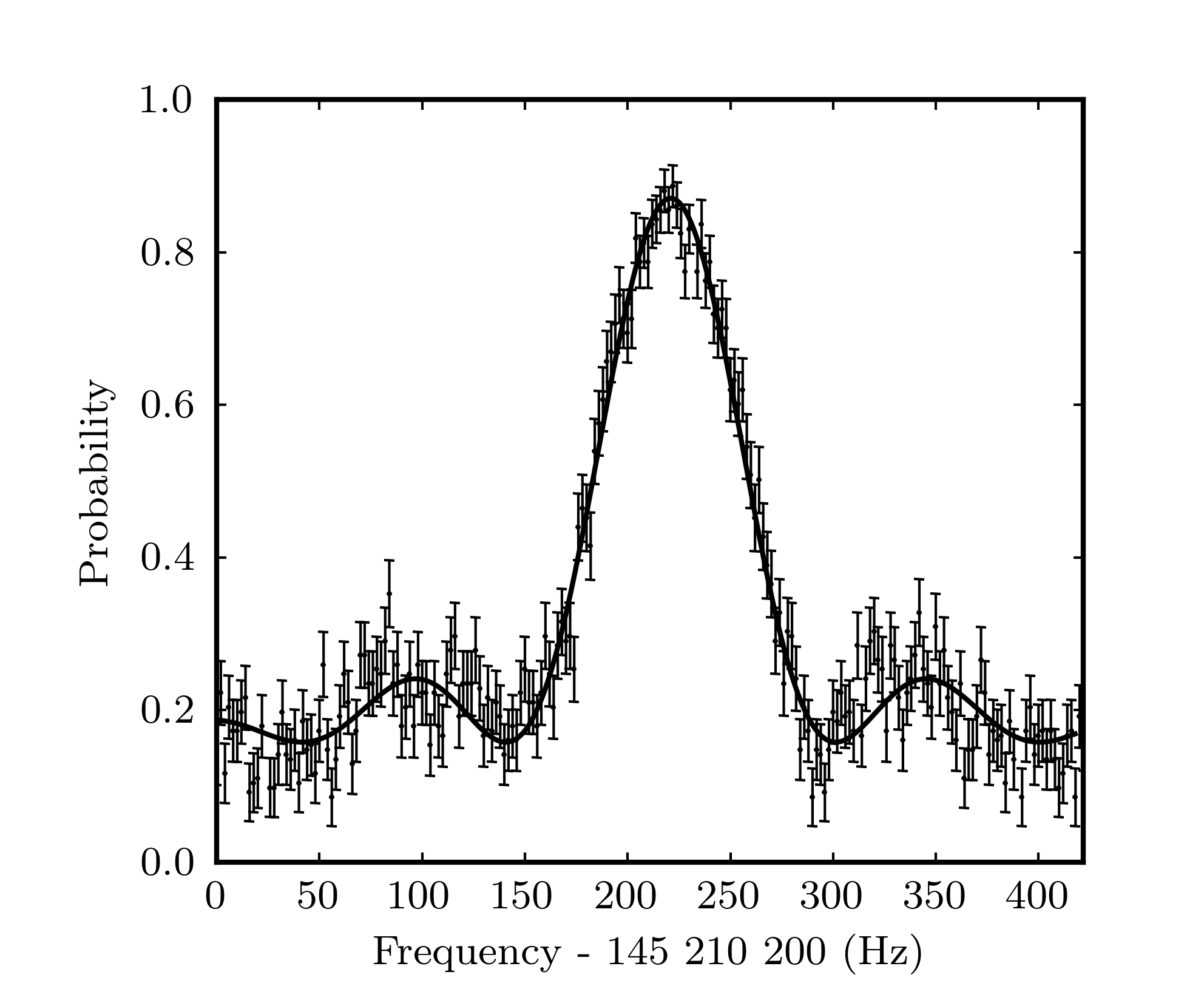}
\caption{Transition probability as a function of the external rf frequency for the $\left|F''=0, m_{F''}=0\right\rangle$$\leftrightarrow$$\left|F''=1, m_{F''}=0\right\rangle$ transition.}
\label{repump}
\end{figure}

An example of a hyperfine splitting measurement for a fixed value of the magnetic field is shown in Fig.~\ref{repump} where the transition probability for $\ket{F''=0,m_F=0}$ to $\ket{F''=1,m_F=0}$ is plotted as a function of the external rf frequency. The rf drive power is adjusted to give an on resonant $\pi$-pulse time of approximately $10$~ms resulting in a resonance width (full width half maximum) of approximately 50~Hz. The resonance is scanned in 2~Hz frequency steps, and the probability of undergoing the hyperfine transition is determined by the average over 200~measurements. The data is fitted \cite{LM} via a $\chi^2$ minimization to a sinc function with additional offset and amplitude parameters to account for imperfect shelving. From the fits we also extract the $68\%$ confidence limits on the fit parameters giving a determination of the center transition frequency with an accuracy of about 1~Hz. At this same fixed magnetic field, a similar measurement is also performed for the $\Delta m_F=\pm 1$ transitions. For this case we use a $\pi$-pulse time of 0.5~ms and scan the resonances in 20~Hz steps. This gives a measurement of the resonant frequency for these transitions with an accuracy of about $20$~Hz which corresponds to a field accuracy of about $50\,\mathrm{\mu G}$.

The time to perform the $\Delta m_F=0$ and field calibration measurements at a particular value of the magnetic field is about 15~min. It is therefore necessary to consider the effects of magnetic field drifts which give rise to additional errors in the field calibration. Hence we have monitored the variation of the magnetic field over the course of a day by measuring the first order Zeeman shift of the $\Delta m_F=+1$ transition. From these measurements we estimate the one standard deviation error in the field calibration to be approximately $450\,\mathrm{\mu G}$ which dominates the $50\,\mathrm{\mu G}$ error extracted from the $\Delta m_F=\pm 1$ field calibration measurements.

\begin{figure}
\includegraphics[width=0.48\textwidth]{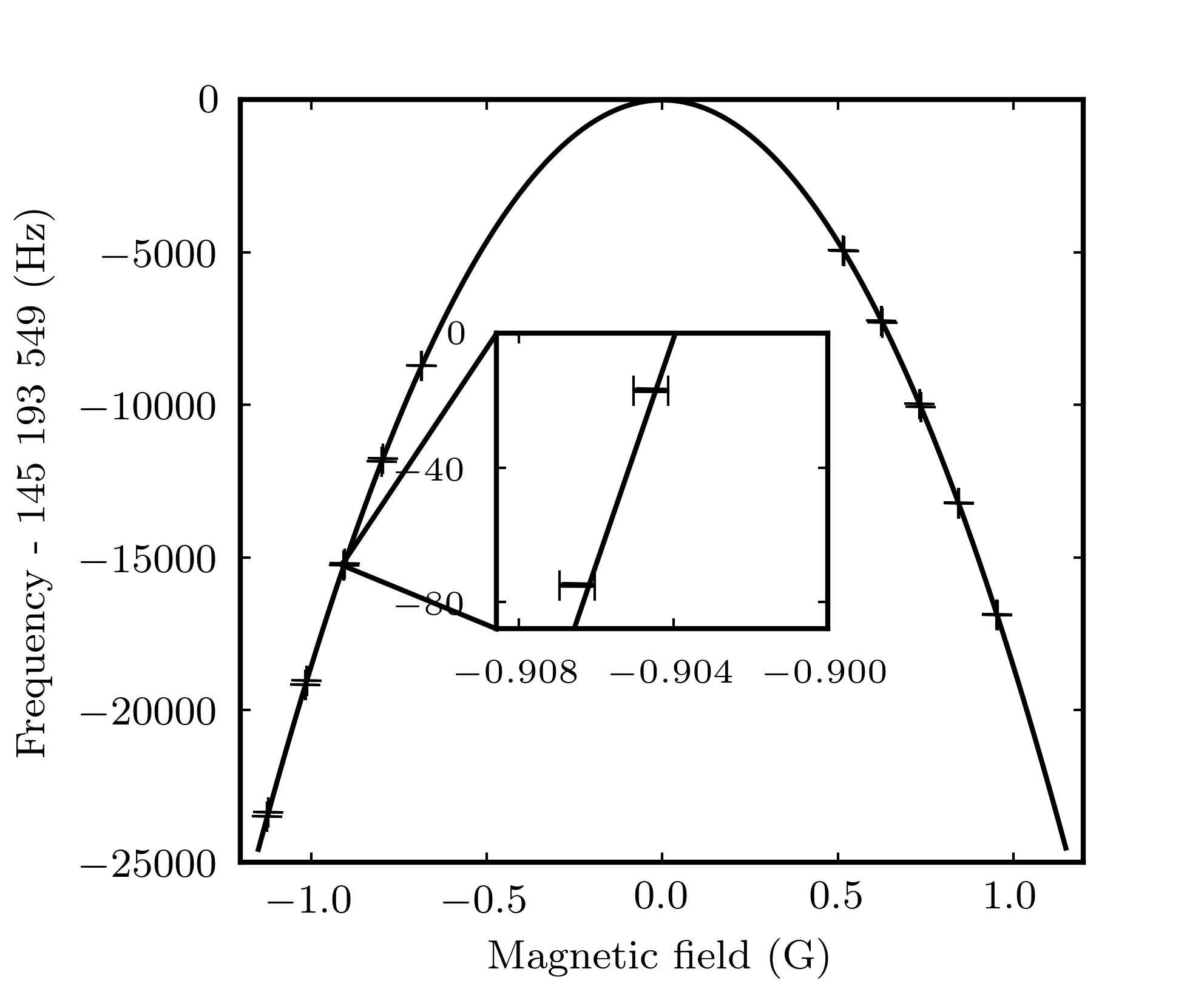}
\caption{Measured hyperfine splitting as a function of the magnetic field for the $\left|F''=0, m_{F''}=0\right\rangle$$\leftrightarrow$$\left|F''=1, m_{F''}=0\right\rangle$ transition.}
\label{splittings}
\end{figure}

Measurements for the hyperfine interval, $\delta W_0$, are shown in Fig.~\ref{splittings}. For each hyperfine interval we have taken two sets of data on separate days. For each set, we measured the transition frequency at $10$ values of the field: 5 with the field above zero and 5 below. The inset highlights two points taken at a similar field setting on separate days. Confidence limits from the fits to the resonance scans determine the vertical errors, which are smaller than the thickness of the lines shown in the inset. We fit the data for each hyperfine interval to a quadratic form, $\alpha_k B^2-\delta W_k$, using a $\chi^2$ minimization. Since the second order Zeeman shift coefficients, $\alpha_k$ are determined by the three values of $\delta W_k$, we fit all three quadratic forms simultaneously. However, we may still determine a $\chi^2$ statistic for each data set. Since the $\alpha_k$ are only weakly dependent on the $\delta W_k$, the minimization procedure is equivalent to three independent single parameter fits with the $\alpha_k$ fixed to values consistent with the fitted values of $\delta W_k$. In Table~\ref{fval} we give the $\delta W_k$ along with the reduced $\chi^2$ for each fit. Errors reported here are again the $68\%$ confidence limits extracted from the fits. For the field calibration and calculation of the $\alpha_k$ we have used values of $g_J$ and $g_I$ reported in \cite{Knoll} and \cite{Marx} respectively.

It is worth noting that the quadratic form used for the fitting is only an approximation. As the field strength increases, higher order terms become important, which can shift the zero field value extracted from a quadratic fit. The smaller the hyperfine splitting, the larger the effect. It is for this reason we have restricted our field values to $\sim 1\,\mathrm{G}$. Over the range of field values we have used, we estimate that this effect could be as large as $\sim 0.5\,\mathrm{Hz}$ for $\delta W_0$. For field values in the range $1\sim 2 \,\mathrm{G}$ this systematic error would be comparable to the uncertainty in our current measurement.

Additional systematic errors could arise from three possible sources: off-resonant coupling of the rf drive to other Zeeman levels, coupling of the ion trap rf field to the ion via micromotion, or AC Stark shifts due to leakage light from the $650\,\mathrm{nm}$ lasers used in the experiment. In Table~\ref{errors} we give estimates for these shifts. For the rf drive field, the largest effect comes from the field calibration measurements since the Rabi rate for these measurements is significantly larger than for the $\Delta m_F=0$ measurements. In this case, the Stark shifts vary with the magnetic field and can alter the quadratic form in Fig.~\ref{splittings} causing a shift in the zero-field point. This effect is largest for the $F=0\leftrightarrow F=1$ transition and the estimate given in the table assumes equal power available for the $\Delta m_F=0,\pm 1$ transitions. The effects of micromotion were estimated using the results in \cite{Wineland_JAP}. Additionally, we shifted the ion position to induce micromotion and saw no measurable shift in the resonance. For the $650\,\mathrm{nm}$ shelving beam, it was found that $1\,\mathrm{\mu W}$ of leakage light was enough to shift the resonances by $\sim 60\,\mathrm{Hz}$ consistent with calculations. The estimate given in Table~\ref{errors} is based on this measurement and the additional attenuation used to remove the shift. For the repumping beams we attempted to measure the leakage light level by measuring the rate at which population was removed from the D$_{3/2}$ level. As there was no statistically significant population loss over $200\,\mathrm{ms}$, the estimate given is an upper bound.

The hyperfine coupling constants inferred from the measured values of $\delta W_k$ are given in table~\ref{hfc}. To account for the possible systematic errors we have added $0.5\,\mathrm{Hz}$ to the errors given in table~\ref{fval}. The correction terms were calculated from the expressions given in \cite{Beloy_pra_012512_2008}, using the matrix elements $\langle 5 D_{3/2}||T_1||5D_{5/2}\rangle=-995(10) \,\mathrm{MHz/\mu_N}$ and $\langle 5 D_{3/2}||T_2||5D_{5/2}\rangle=339(5)\,\mathrm{MHz/b}$ \cite{Sahoo_prep}. The magnetic moment $0.937365(20)\mu_\mathrm{N}$ and quadrupole moments $246(1)\mathrm{b}$ where taken from \cite{Werth_dipole} and \cite{bijaya_ba+_atom_quad} respectively.

The final values obtained for $A$ and $B$ are within three standard deviations of the results reported in \cite{Hove} with a 30-fold reduction in the uncertainty. In addition, our value for $C$ has an accuracy of $0.2\%$ and is the first observation of the magnetic octupole moment in $^{137}\mathrm{Ba}^+$. Using Eq.~(\ref{omega}) we calculate an octupole moment of
\begin{displaymath}
\Omega(^{137}\mathrm{Ba}^{+})=-0.06290(67)~(\mu_\mathrm{N} \times \rm{b}).
\end{displaymath}
This, to our knowledge, is the most accurately determined octupole moment to date.

\begin{table}[ht]
\caption{\label{fval}Measured hyperfine intervals, $\delta W_k$, for the 5D$_{3/2}$ manifold of $^{137}$Ba$^{+}$.}
\begin{ruledtabular}
\begin{tabular}{lll}
Transition & $-\delta W$ (Hz) & Reduced $\chi^2$ \\
\hline F=0 $\rightarrow$ F=1 & 145 193 549.3 (2.8) & 1.01 \\
F=1 $\rightarrow$ F=2 & 334 921 347.13 (89) & 1.60 \\
F=2 $\rightarrow$ F=3 & 613 730 628.08 (22) & 0.47 \\
\end{tabular}
\end{ruledtabular}
\end{table}
\begin{table}[ht]
\caption{\label{errors}Estimates of the possible systematic errors.}
\begin{ruledtabular}
\begin{tabular}{ll}
Source & Error estimation \\
\hline Off resonant rf coupling & $\sim 0.1\,\mathrm{Hz}$ \\
Micromotion & $\sim 5\,\mathrm{mHz}$\\
Stray Raman light & $\sim 60\,\mathrm{mHz}$ \\
Stray repump light & $\lesssim 0.5\,\mathrm{Hz}$ \\
Higher order terms ($\delta W_0$) & $\sim 0.5\mathrm{Hz}$\\
\end{tabular}
\end{ruledtabular}
\end{table}
\begin{table}[ht]
\caption{\label{hfc}Hyperfine coupling constants.}
\begin{ruledtabular}
\begin{tabular}{lrrr}
              & A (Hz)           & B (Hz)          & C (Hz) \\
\hline Uncorr.       & 189730524.90(31) & 44538793.6(1.0) & 32.465(42) \\
$\eta$ corr.  &          805(16) &       $-$1610(32) & $-$ \\
$\zeta$ corr. &      $-$228.5(4.2) &        $-$571(10) & 4.081(75) \\
Corr.         &    189731101(17) &    44536612(34) & 36.546(86)\\
\end{tabular}
\end{ruledtabular}
\end{table}

In conclusion we have performed high precision measurements of the hyperfine structure in $^{137}\mathrm{Ba}^+$. Our measurements have greatly reduced the uncertainty of the currently available hyperfine structure constants and have provided an estimate of the nuclear octupole moment, $\Omega$, accurate to $1\%$. We note that our inferred value of $\Omega$ has the opposite sign compared to the estimate given in \cite{Beloy_pra_052503_2008}. This estimate is based on the shell model for which such discrepancies have also been reported for Cesium \cite{Gerginov_prl_2003} and Rubidium \cite{Gerginov_cjp}. Thus our measurements may present an interesting test case for comparison with nuclear-structure calculations. Approximately $10\%$ of the estimated value for $\Omega$ is given by the correction factor, $\zeta$. Future planned measurements on the 5D$_{5/2}$ manifold of the same ion will allow the dependence on this factor to be eliminated. This would also provide additional consistency checks between measurements and calculations.

\begin{acknowledgments}
This research was supported by the National Research Foundation and the Ministry of Education of Singapore. A part of the calculations were carried out using 3TFLOP HPC Cluster at PRL, Ahmedabad.
\end{acknowledgments}

\providecommand{\noopsort}[1]{}\providecommand{\singleletter}[1]{#1}%

\end{document}